\documentclass[12pt,a4paper,reqno]{amsart}
\usepackage{amsmath,mathtools,cancel,empheq}
    \usepackage{amssymb}
    \usepackage{amsthm}
    \usepackage{amsfonts}
    \usepackage{braket}
    \usepackage{psfrag,color}
    \usepackage{bbold}

\begin{document}
{\bf Non-Particulate Quantum States of the Electromagnetic Field in Expanding Space-Time}

\bigskip
\textit{{P. Broadbridge},
{S. Becirevic}
 and {D. Hoxley,}}\\
 Dept of Mathematical and Physical Sciences,\\
 La Trobe University,
 Bundoora VIC 3086, Australia.

\bigskip

\textit{Abstract}\\
\footnotesize{A quantum field has been coupled to a space-time with accelerating expansion. Dynamical modes are destabilised successively at shorter material wavelengths as they metamorphose from oscillators to repellers. Due to degeneracy of energy levels, the number of unstable modes increases at an accelerating rate, sufficient to account for a significant proportion of cosmic energy. For the subsystem spanned by a finite basis of unstable runaway modes, the quantum Hamiltonian is unbounded below. There is no Bogoliubov transformation by which that subsystem Hamiltonian can be expressed as a linear combination of number operators. For the remaining subsystem spanned by an infinite number of oscillator modes, there is an appropriate vacuum state in a Fock-Cook representation of the field algebra.
The massless quantum vector field of electromagnetism is considered when it is minimally or more generally coupled to an expanding space-time. For a significant class of models, including minimal coupling models and the exponential  de Sitter universe coupled to the Ricci curvature tensor, the field equations are equivalent to the Proca equation with time-dependent mass.}

\bigskip

\textit{keywords:} % use a minimum of 5 kwrds, separate them with a comma
vacuum instability, electromagnetic field, FLRW, number operators, dark energy.

\normalsize
\section{Introduction} % first section MUST be titled Introduction, and feature introductory text; do NOT change this title

The simplest Bosons that we encounter are phonons that mediate the equally spaced energy levels of a quantised harmonic oscillator. In general, small oscillations of a many-body system about a stable equilibrium state can be approximated as a Hamiltonian system with positive definite quadratic Hamiltonian.  It is well known that for such a system, there exists a linear canonical (or symplectic) transformation on phase space by which the Hamiltonian is seen to be that of a system of independent harmonic oscillators, with angular frequencies $\omega_j;~j=1\cdots N$, that are the frequencies of the elementary excitations of the system with $N$ degrees of freedom. That concept can be extended to infinite degrees of freedom in the guise of field theories.
Relativistic free field theories are generally based on invariant linear wave equations that involve well defined particle mass and spin that are invariants of the inhomogeneous Lorentz group. They can still be represented mathematically as a system of independent harmonic oscillators. In the early years of quantum field theory, that approach was developed systematically for the electromagnetic field as well as for a massive scalar field \cite{Pauli}. Each component oscillator can be quantised and each has its own number operator that commutes with the quantum Hamiltonian. This leads to the Fock representation of a quantum field in which eigenstates of the Hamiltonian are labelled by particle occupation numbers.  \\
Because in reality we deal mainly with stable systems whose Hamiltonians have a lower bound, it is not so widely known that for unstable systems, there does not even exist a complete set of number operators that commute with the Hamiltonian. Under the action of the group of symplectic transformations, indefinite quadratic Hamiltonians have five types of irreducible canonical form beyond that of the harmonic oscillator. For the other classes, there is no vacuum state that can be defined as a zero-particle state. The Fock representation breaks down. The Hamiltonian has continuous spectrum, unbounded below, and a basis of solutions to the governing wave equation must have runaway non-oscillatory modes. \\
 Minkowski space is only a local approximation. Boson fields must be coupled to the gravitational field of a universe that is expanding and accelerating. The fields may become unstable. When a scalar Klein-Gordon field is weakly coupled to the gravitational field of a spatially flat and exponentially stretching de Sitter space-time, the component building blocks of one dynamical degree of freedom are harmonic oscillators with time-dependent decreasing force constant that eventually will be negative. They may transition from being an attractive simple oscillator to a repulsive simple repeller. With transitions to instability occurring over time at successively shorter material wavelengths, the basic components of the field include a finite number of simple repellers that have pure imaginary angular frequencies. At a sequence of finite times, each solution mode ceases to oscillate.  Working through the steps of a long-established and trusted approach \cite{Pauli}, the quantum Hamiltonian consists of two parts; one is equivalent to the familiar infinite sum of commuting number operators, while the other is equivalent to a finite sum of pair creation operators and pair annihilation operators that cannot be transformed to number operators by a Bogoliubov transformation \cite{Deutscher}. The energy eigenstates of the repellers are neither wave-like nor particle-like. They would not resonate with wave detectors or particle detectors. In that sense they are dark.\\
According to prevailing theoretical models and observations, currently the energy of the universe is predominantly dark energy, associated with exponential expansion. In its previous phase, cosmological mass-energy was predominantly due to matter, leading to an expansion that was of the $\frac 23$ degree in cosmic time. Until the end of the plasma phase, around 360,000 years after the big bang, most of the energy budget was filled by photon and neutrino radiation, leading to expansion in proportion to square root of time.
\\Due to a period of rapid inflation, ending around around 10$^{-24}$s after the big bang, much of the universe is beyond the horizon of light signals. Inflation explains why distant  parts of space that have recently come into view, are still in thermal equilibrium with the microwave background. Because of the variable expansion rates, we are motivated to ask which expansion rate laws cause destabilisation of scalar and electromagnetic vector fields when they are coupled to Friedmann-LeM\^aitre-Robertson-Walker (FLRW) isotropic homogeneous expanding space-times. Following on from that, we plan to quantise the destabilised electromagnetic field.\\
For any FLRW expansion law, one may trace sufficiently far back in time to a field Hamiltonian that did commute with a complete set of number operators, and there was a well defined vacuum state of lowest energy. Having defined such a state, one may propagate it forwards in time. That is the general idea behind the Bunch-Davies vacuum \cite{Bunch}. For gentle expansion accelerations, the vacuum states at one time may have particles, as counted by number operators that commuted with the Hamiltonian at an earlier time. A basic example of relativity of particle number is given by the Unruh effect \cite{Unruh}, in which the temperature of the particle spectrum is proportional to the constant acceleration of an observer. For some less gentle but reasonable expansion laws, the Bogoliubov transformation that relates different sets of creation and annihilation operators for single-particle basis wave functions appropriate at different times, will not be implementable by a unitary transformation. In those cases, the two Fock representations will not be physically equivalent and the energy of the vacuum may be infinite, as measured with the Hamiltonian of an earlier time. Forty  years ago that phenomenon was investigated in reference \cite{Kodama}. Due to the degeneracy of energy spectral values of such symmetric fields, the number of unstable modes increases rapidly as the square of the spatial wave number of the destabilising mode \cite{Kodama}. It is then feasible that the energy density of the unstable modes may be approximately constant at large times \cite{Deutscher}.\\
Beyond asking whether vacuum states at two different times are related, the more fundamental problem that we are addressing here, is that the evolving Hamiltonian may eventually generate unstable field dynamics, in which case the Fock representation breaks down and the very existence of such a vacuum state is negated, at least for a subsystem of finite degrees of freedom. 

\section{Symplectic Transformations and Bogoliubov Transformations}
The relationship between symplectic transformations and Bogoliubov transformations was reviewed and developed in reference \cite{PBThesis}. Most dynamical systems in mathematical physics can be derived from a principle of least action,
$$\delta \int _a^b L({\bf q}; \dot {\bf q})dt=0,$$
where $L$ is the Lagrangian. 
 If $\{q_j:~j=1\cdots N\}$ is a general coordinate system for configuration space, the conjugate momentum variables are $p_j=\partial L/\partial \dot q_j.$ Note that the conjugate pairs $(q_j,p_j)$ need not have the standard measurement dimensions of displacement and lineal momentum. However prior to non-dimensionalisation, $q_jp_j$ should have the standard dimensions of action. Hamiltonian mechanics allows additional generalisation. Although $(q_j,p_k)$ must represent phase space, $q_j$ need not represent configuration space. For example, stable systems should be transformable to a system that is invariant under Born reciprocity \cite{Born}, $(Q_j,P_j)=(-p_j,q_j)$.
Hamilton's equations of motion in phase space are 
\begin{eqnarray}
\dot q_j=\frac{\partial H}{\partial p_j}\\
\dot p_j=\frac{-\partial H}{\partial q_j};~~j=1\cdots N,
\end{eqnarray}
where $H({\bf q};{\bf p})=\sum_1^N p_j \dot q_j.$
As set out in reference \cite{Whittaker}, an equilibrium point of that system may be shifted to the origin and the Hamiltonian is then approximately quadratic $H=\frac 12{ \bf z}^t\hat H {\bf z}+\mathcal O(|{\bf z}|^2),$ where ${\bf z}=({\bf q,p})^t$ and $\hat H$ is a real symmetric $2N \times 2N$ matrix. Hamilton's equations are preserved by canonical transformations $(q_j,p_j)\mapsto (Q_j({\bf z}),P_j({\bf z}))$ that preserve the Poisson bracket relations 
\begin{eqnarray}
\label{PB}
\{Q_j,Q_k\}=\sum_{m=1}^N \frac{\partial Q_j}{\partial q_m} \frac{\partial Q_k}{\partial p_m}-\frac{\partial Q_j}{\partial p_m} \frac{\partial Q_k}{\partial q_m}=0,\\
\{P_j,P_k\}=0,\\
\{Q_j,P_k\}=\delta_{j,k}.
\end{eqnarray}
Linear canonical transformations ${\bf z}=C{\bf Z}$, are represented by real $2N\times2N$  symplectic matrices that preserve a non-degenerate skew-symmetric form with Gram matrix G.  $C^t G C=G$ with $G^t=-G$. Without loss of generality, 
$G=\begin{pmatrix}
         0                                    &-I_N  \\
        
        I_N                                        &0\\
        
\end{pmatrix}$.
Then the quadratic forms transform as $\hat H'=C^t\hat H C\iff G\hat H'=C^{-1}G\hat HC.$ It is well known that a positive definite quadratic Hamiltonian can always be transformed to that of a system of independent harmonic oscillators with angular frequencies $\omega_j$ that are real eigenvalues of $iG\hat H$. Since $iG\hat H$ transforms by a similarity transformation, these angular frequencies are invariants.   Among the canonical forms $G\hat H$ for conjugacy classes of generators  of the symplectic group, there are five different types \cite{Williamson}. Their irreducible Jordan blocks may be arbitrarily large and they account for complex frequencies. A simple repeller with $N=1$ and $H=\frac 12 p^2-\frac 12 \alpha^2 q^2$ has pure imaginary angular frequencies $\pm i\alpha$ so it cannot be transformed to a simple oscillator that has real angular frequencies.
 
\subsection{Heisenberg Algebra}
In quantum mechanics, real functions on phase space are replaced by self-adjoint operators. Under the Dirac correspondence principle, the skew-symmetric commutator brackets between quantum observables $q_j$ and $p_k$ take the same values as for the Poisson brackets in (\ref{PB}) multiplying $i\hbar I$ where $2\pi \hbar$ is Planck's constant. These are the canonical commutation relations (CCR). Those commutator relations generate the Heisenberg Lie algebra and the corresponding unitary operators $e^{i({\bf q,p})^t{\bf  z}};~{\bf z}\in \mathbb R^{2N}$, are the building blocks of the Heisenberg-Weyl group.\\
Now it is convenient to rescale quantum phase space variables to non-dimensional form, choosing scales $m_s, \ell_s$ and $t_s$ for mass, length and time so that $\hbar=1$. 
Conforming with reference \cite{Deutscher}, we choose $\ell_s=R_0$ (universe radius at present time as reference), $t_s=R_0/c$ and $m_s=\hbar/cR_0$. This also scales $c$ to 1, so that $x_0$ may be identified with $t$. When dealing with electromagnetic phenomena, there is one more independent quantity, such as charge, to be scaled. When dealing with free fields, it is more convenient to choose the electrical permittivity of the vacuum $\mu_0$ as the permittivity scale. Since $c^2=\mu_0\varepsilon_0$ is scaled to 1, the magnetic permeability $\mu_0$ automatically scales to 1 also. 
\subsection{Boson Algebra}
After non-dimensionalising, let $a_j=2^{-1/2}(q_j+ip_j)$. Then $a_j$ and their adjoints $a_j^\dagger$ satisfy the Boson CCR: $[a_j,a_k]=0,~~[a_j^\dagger,a_k^\dagger]=0,~~[a_j,a_k^\dagger]=\delta_{jk}$. Hence, quadratic Hamiltonians may be expressed as quadratic combinations of annihilation and creation operators. With 
unitary matrix $\mathcal M=2^{-1/2}\begin{pmatrix}
         I_N                                    &I_N  \\
        
        -iI_N                                        &iI_N\\
        
\end{pmatrix}$,
\begin{equation}
\label{equiv}
H=\frac 12 {\bf z}^t\hat H {\bf z}=\frac 12({\bf a^\dagger,a})\mathcal D ({\bf a,a^\dagger})^t,
\end{equation} 
where $\mathcal D=\mathcal M^\dagger \hat H\mathcal M$. Now $\hat I\mathcal M^\dagger=-i\mathcal M^\dagger G$, where $\hat I = diag[I_N,-I_N]$. It follows that $\hat I\mathcal D=\mathcal M^{-1}(-iG\hat H)\mathcal M$. By similarity, $\hat I\mathcal D$ must have the same eigenvalues as $-iG\hat H$, namely the characteristic frequencies $\pm\omega_j$. For example, the Hamiltonian of a harmonic oscillator may have $H=\frac 12 P^2+\frac 12 \omega^2 Q^2.$ After the canonical transformation $(Q,P)=(\omega^{-1/2}q,\omega^{1/2}p)$,\\
 $H=\frac 12 \omega P^2+\frac 12 \omega Q^2=\frac 12 \omega a^\dagger a+\frac 12 \omega aa^\dagger=\frac 12 \omega +\omega a^\dagger a.$ Whenever $\hat H$ (and therefore $\mathcal D$) is positive definite, the Hamiltonian may be expressed as a ground state energy plus a linear combination of commuting number operators $\sum_j\omega_j a_j^\dagger a_j$. This may require a canonical transformation or a Bogoliubov transformation to render the Hamiltonian in this standard form. A Bogoliubov transformation $(a_j,a_k^\dagger)\mapsto (b_j,b_k^\dagger)$ involves  linear combinations of annihilation and creation operators that preserve the Boson CCR as well as Hermitian conjugacy between $b_j$ and $b_j^\dagger .$ In order to maintain the CCR, the linear transformation $\begin{pmatrix} {\bf a} \\{\bf a^\dagger}\end{pmatrix}=\mathcal T\begin{pmatrix} {\bf b} \\{\bf b^\dagger}\end{pmatrix}$ satisfies $\mathcal T^\dagger \hat I\mathcal T=\hat I$; i.e. it must be pseudo-unitary in $U(N,N)$. To satisfy the  Hermitian conjugacy, it must have the block structure 
 \begin{equation}
 \label{BlockBog}
 \mathcal T=\begin{pmatrix} T_1 &T_2\\ T_2^* & T_1^* \end{pmatrix}.
 \end{equation}
  Because of the equivalence (\ref{equiv}), the group of Bogoliubov transformations is isomorphic to the real symplectic group. It must preserve characteristic frequencies. For indefinite Hamiltonians, the characteristic frequencies need not be real. For example the indefinite self-adjoint Hamiltonian $H=\frac 12 [i\alpha aa-i\alpha a^\dagger a^\dagger]$ has pure imaginary characteristic frequencies $\pm i\alpha$. It cannot be transformed to a linear combination of number operators which would have real frequencies. In that case, there can be no number operators that commute with the indefinite Hamiltonian \cite{Chaiken}. There is no zero-particle vacuum state that would have any particular significance for that dynamical system. Energy eigenstates of the quantum Hamiltonian could not be labelled by particle numbers.
\subsection{Indefinite Metric on Single Mode Space}
If $iG\hat H$ has real eigenvalues $\pm \omega_k$, then there exists a canonical transformation that reduces the system to a collection of independent harmonic oscillators.  The general solution of Hamilton's equations for a single harmonic oscillator with $H=\frac 12 (p^2+\omega^2 q^2)$ is 
\begin{equation}
q=Ae^{-i\omega t}+A^*e^{i\omega t}; ~p=-i\omega Ae^{-i\omega t}+i\omega A^*e^{i\omega t}.
\end{equation}
After quantisation, the amplitudes must be generalised to operators $A$ and $A^\dagger$. It is well known that scaled amplitudes $A/\sqrt{2\omega}$ and $A^\dagger/\sqrt{2\omega}$ satisfy the Boson CCR for annihilation and creation operators. The classical solution space may be regarded as a single-particle space. This is behind the general idea of second quantisation; in that case, the wave function solution space for the classical or 'first quantised' field is the single particle space for the second-quantised field. In the Cook construction of a Fock representation, n-particle space is a n-fold tensor product of the classical solution space \cite{Cook}. However for unstable systems with quadratic Hamiltonian, the CCR among operators $q$ and $p$ are incompatible with the Boson CCR among quantised amplitude operators $A$ and $A^\dagger$. For example, consider a charged repeller with $H= p^\dagger p-q^\dagger q=\frac 12[(p_1^2+p_2^2)-(q_1^2+q_2^2)]$,
where $q=2^{-1/2}(q_1-i q_2)$ and $p=2^{-1/2}(p_1+ip_2)$. The general solution is of the form
\begin{equation}
q=\sqrt 2 Ae^{-t}+\sqrt 2 B e^t;~~p=-\sqrt 2 A^\dagger e^{-t}+\sqrt 2 B^\dagger e^t.
\end{equation}

If we demand that the mode amplitude operators satisfy the Boson CCR, $[A,A^\dagger]=I,~[B,B^\dagger]=1$ and other commutators vanish, then it follows that 
\begin{equation}
[q_1,q_2]=-2i\cosh (2t).
\end{equation}
Similar commutators appear among field operators $\Phi(x^\mu)$ and $\Phi(y^\mu)$ at space-like separation $x^\mu-y^\mu$ when the field obeys a Klein-Gordon equation with $m^2<0$. Such a field may be decomposed as a system of independent oscillators and independent repellers. In reference \cite{Schroer}, the imaginary-mass Klein-Gordon equation was quantised using an approach that is opposite to the above. If one assumes standard CCR among the operators $q$ and $p$, then it follows that 
\begin{equation}
[A,B^\dagger]=\frac 12 i.
\end{equation}
whereas other commutators vanish. If one assumes the existence of a normalised cyclic vector  $\psi_0$ in a Hilbert space, satisfying $A\psi_0=0$ and $B\psi_0=0$, then there follows the inner product $\braket {A^\dagger \psi_0|B^\dagger \psi_0}=\frac 12 i$. In the basis $\{A^\dagger \psi_0, B^\dagger \psi_0\}$, single-mode complex Hilbert space has a sesquilinear inner product with Gram matrix $\begin{pmatrix} 0 &i/2\\ -i/2^* & 0 \end{pmatrix}$, which is indefinite with eigenvalues $\pm \frac 12$. An indefinite inner product arose in the Gupta-Bleuler method to quantise the electromagnetic field \cite{Kallen}. However in flat Minkowsi space, the dynamics maintain a separation between physical states of positive norm and unphysical states of negative and zero norm, so that the latter may be factored out. This occurs when the basic components are of the form $ H=\frac 12 (p_1^2+\omega_1^2 q_1^2)-\frac 12 (p_2^2+\omega_2^2 q_2^2)$. Then the component with positive definite energy commutes with the total Hamiltonian. In all other cases of indefinite quadratic Hamiltonian, this is no longer true \cite{PB1983}.
\section{Electromagnetic Field}
\subsection{From Finite to Infinite Degrees of Freedom}
Field variables $\Phi (x,t)$ take time-dependent values at each point in space. Therefore they have infinite degrees of freedom, often labelled by space location $x$. Most of the fields of mathematical physics obey an action principle
$$\delta \int \int _\Omega \mathcal L(\Phi(x,t);~\Phi,_\mu) dV~dt~=0,$$
where the Lagrangian density $\mathcal L$ is meant to depend on $\Phi$ as well as its time derivative $\Phi,_0=\partial\Phi/\partial x^0$ and space derivatives $\Phi,_j=\partial\Phi/\partial x^j;~j=1,2,3.$ Here $\Omega$ is the spatial domain. The canonical conjugate field is $\Pi (x,t)=\partial\mathcal L/\partial \Phi,_0$. Provided $\mathcal L$ depends on $\Phi,_0$, canonical quantisation may proceed in the usual way. Equal-time canonical commutation relations now extend the Kronecker delta to Dirac delta function: $[\Phi(x,t),\Pi(y,t)]=i\delta({\bf x}-{\bf y}).$ 
The dynamical field equations are the Euler-Lagrange equations that follow from the action principle:

$$\frac{\partial\mathcal L}{\partial\Phi}-\partial_\lambda \frac{\partial\mathcal L}{\Phi,_\lambda}=0.$$
In relativistic formulations we will assume the Einstein summation convention that Greek indices repeated at lower (covariant) and upper (contravariant) levels, will be summed from 0 to 3, whereas repeated Roman indices will be summed from 1 to 3. It will be assumed that space-time is a pseudo-Riemannian manifold with covariant symmetric metric tensor $g_{\mu\nu}$ with metric defined locally by $(\delta s)^2=g_{\mu\nu}dx^\mu dx^\nu.$  Contravariant components $g^{\mu\nu}$ of the metric tensor are defined to be the matrix entries of the inverse of $g_{\mu\nu};~~g^{\mu\nu}g_{\nu\lambda}=\delta^\mu_{~\lambda}$.\\
Having obtained a basis $\{f_k\}$ of solutions $f_k(x,t)$ of the field equations, we may re-express the quantised Hamiltonian  in terms of operators $a_k$ that satisfy Boson CCR. For fields in Minkowski space the labels $k$ may be wave vectors in the Fourier basis, such that all Boson CCR including $[a_{\bf k},a_{\bf p}^\dagger]=\delta({\bf k}-{\bf p})$, are satisfied. On bounded domains, the basis is countable and the commutators are zero or Kronecker delta.\\
 Just as in quantum mechanics with finite degrees of freedom, field theories may be described in terms of self-adjoint quantum fields \cite{Friedrichs}, or alternatively in terms of creation and annihilation operators \cite{Berezin}. A Bogoliubov transformation is implementable as a unitary transformation $b_k=Ua_k$ in a Hilbert  space representation of the quantum field, if and only if the off-diagonal creation-to-annihilation block $T_2$ in (\ref{BlockBog}), is of the Hilbert-Schmidt class, i.e. trace of $T_1^\dagger T_1$ is finite. Many classical field models are Hamiltonian, leaving invariant some symplectic form between pairs of solutions \cite{Segal}. An infinite dimensional symplectic transformation C is unitarily implementable if and only if $C^tC-I$ is of the Hilbert-Schmidt class \cite{Friedrichs}. These two Hilbert-Schmidt conditions are equivalent \cite{PBThesis}. The one-parameter symplectic flow given by unstable field dynamics, need not be unitarily implementable. Then in the Fock representation, the vacuum state for a set of evolving annihilation operators may diverge in the norm. The rate of change of energy of an evolving vacuum state may diverge \cite{Hove}. In the context of scalar Boson fields coupled to a FLRW space-time, this problem can be avoided only through conformal coupling \cite{Kodama}.

\subsection{Electromagnetic 4-Potential}
The free-field Maxwell field equations are encapsulated as the zero divergence of the Faraday electromagnetic tensor, along with the Bianchi identity:
\begin{eqnarray}
\nabla_\nu F^{\mu\nu}=0,\\
\partial_\mu F_{\alpha \beta}+\partial_\alpha F_{\beta \mu}+\partial_\beta F_{\mu\alpha}=0.
\end{eqnarray}  
Here, $\nabla_\mu$ is the covariant gradient. 
\subsubsection{Contravariant Components and Physical Components}
Consider a vector ${\bf A}=A^\mu {\bf e}_\mu$ in terms of a non-normalised basis in an indefinite inner product space. Let $g_{\mu\nu}={\bf e}_{\mu}\cdot{\bf e}_\nu$.  The `physical' components of a vector are traditionally taken with respect to a normalised basis ${\bf e'}_{\mu}={\bf e}_{\mu}/\sqrt{|g_{\mu\mu}|}$. Now ${\bf A}=A^{'\mu}{\bf e}'_{\mu}$ with $A^{'\mu}=\sqrt{|g_{\mu\mu}|}A^\mu$.
Then ${\bf A}\cdot{\bf A}=A^\mu A^\nu g_{\mu\nu}=A'^\mu \eta_{\mu\nu}A'^\nu$, where $\eta_{\mu\nu}$ is the normalised metric tensor $g_{\mu\nu}/\sqrt{|g_{\mu\mu}g_{\nu\nu}|}$. In the current application, this is just the Minkowski metric.\\
Similarly, components of the tensor $F^{\mu\nu}$ may be normalised to physical components $F'^{\mu\nu}=F^{\mu\nu}\sqrt{|g_{\mu\mu}g_{\nu\nu}|}$.

The matrix entries of skew-symmetric $F'^{\mu\nu}$ are components of electric and magnetic fields:
$F'^{\mu\nu}=\begin{pmatrix}
0 & -E^1 & -E^2 & -E^3\\
E^1 & 0 & -B^3 &B^2\\
E^2 & B^3 & 0 & -B^1\\
E^3 & -B^2 & B^1\ & 0\\
\end{pmatrix}'.$\\
The field equations are also written in terms of the electromagnetic 4-vector potential so that $F_{\mu\nu}=\partial_\mu A_\nu -\partial_\nu A_\mu.$ Because of the skew symmetry of that expression,  
if the space is torsionless, then the covariant derivatives are not required in this equation as the connection terms cancel.\\
It is easily verified that 
\begin{equation}
\label{usualL}
-\frac 14F^{\mu\nu}F_{\mu\nu}=-\frac 14 F'^{\mu\nu}\eta_{\mu\alpha}\eta_{\nu\beta}F'^{\alpha\beta}=\frac 12\sum_j E'^jE'^j-B'^jB'^J.
\end{equation}
This is the usual Lagrangian density of classical electrodynamics.
There may be different choices of gauge by which different potential fields $A_\nu(x,t)$ lead to the same physical fields $E_j$ and $B_j$. The simplest type of wave equation for $A_\mu$ is sometimes satisfied after choosing the Lorenz gauge (a particular form of Lorentz gauge), 
\begin{equation}
\label{Lorenz}
\nabla^\mu A_\mu=0.
\end{equation}
 Then in flat Minkowski space, the free field equations $\partial_\nu F^{\mu\nu}=0$ reduce to the linear wave equation $\partial_\mu\partial^\mu A_\nu=\partial_0^2A_\nu-\nabla^2 A_\nu=0.$ In extensions of the field equations to a curved pseudo-Riemannian manifold, it is natural to replace the 4D Laplacian by the Laplace-Beltrami operator. However it is not known exactly how the electromagnetic 
field couples to the curvature. When the covariant derivatives are allowed to operate on general tensor fields, $\nabla_\mu$ and $\nabla_\nu$ do not necessarily commute. In particular, $[\nabla_\mu,\nabla_\nu]A^\mu=R_{\nu\alpha}A^{\alpha}$, where $R_{\mu\nu}$ is the symmetric Ricci tensor. The consequent ambiguity in extending the flat-space Laplacian is covered by allowing such coupling terms to be forcing terms in the field equations \cite{Li}. If, as in the theory of scalar fields, one also considers coupling to the Ricci scalar $R$, then  the field equations are of the form
\begin{equation}
\label{coupledRA}
\frac{1}{\sqrt {|g|}}\partial_\mu\left(\sqrt{(|g|)}g^{\mu\lambda}\partial_\lambda A^\nu\right)=\zeta g^{\nu\lambda}R_{\alpha\lambda}A^\alpha +\xi RA^\nu,
\end{equation}
where $g=det\{g_{\mu\nu}\}$. This breaks gauge invariance when $a(t)$ is not static. However we use it as an indicative model to examine qualitative consequences of electromagnetic-gravitational interactions. Some classical fields of this type were developed in reference \cite{Cabral}. We show below, that even with minimal coupling, the model wave equations are related to the the flat-space Proca equation with time dependent mass term. For the standard Proca system with mass, the Lorenz gauge condition is not only a convenient choice for simplifying the equations but it is a necessary consequence of Proca's covariant formulation \cite{Zamani}.
In dimensionless terms, the Lagrangian density is of the form
$$\mathcal L=\frac{-1}{4}F_{\alpha\beta}F^{\alpha\beta} \sqrt{(|g|)}+\frac 12\zeta R_{\alpha\beta}A^\alpha A^\beta +\frac 12 \xi R A^\alpha A_\alpha.$$
 In the case of minimal coupling ($\zeta=0$ and $\xi=0$), the theory is gauge invariant since the Lagrangian density is expressible in terms of $F_{\mu\nu}$. 
Otherwise, the theory is not gauge-invariant over cosmological scales and the 4-potential will be viewed as a measurable field. However numerical values of the coupling constants are not known. Another popular model is conformal coupling $\xi=\frac 16$ and $\zeta=0$. Independently of the form of $a(t)$, that allows the field equations to be transformed to the standard vector-valued wave equation by a change of variables. However we show here that a special case of the expansion factor, namely the exponential case that our universe is approaching, allows other conformal combinations 
\begin{equation}
\xi=\frac 16-\frac\zeta 4.
\end{equation}
This includes the simple case $\xi=0$ and $\zeta=\frac 23$.
\subsection{Restriction to Spatially Flat FLRW Universes}
Apart from relative deviations of less than 10$^{-4}$  observed in cosmic microwave background temperature, at cosmological scales the universe is approximately homogeneous and isotropic, with flat spatial cross sections. Since the 1970s there has been considerable development of the quantum theory of scalar fields coupled to FLRW expanding space-time \cite{Birrell, Mukhanov}.

When analysing dynamical fields, we choose to use the convention that time-like separations are positive. The homogeneous isotropic metric is then of the form 
$$(ds)^2=(dx^0)^{2}-a(x^0)^2\left[(dx^1)^{2}+d(x^2)^{2}+(dx^3)^{2}\right]$$
Here, $x^j;~~j=1\cdots 3$, are material coordinates that remain constant on a fixed material point that is moving with the average cosmological expansion. $t$ is cosmic time as measured by an observer attached to such a material point. $a(t)$ is the expansion factor for stretching space. In the FLRW form of (\ref{coupledRA}), we need exact expressions for the Ricci scalar and the Ricci tensor. These are
\begin{eqnarray}
R_{\mu\nu}=\hbox{diag}\left[-3\frac{a"(t)}{a}, ~2(a')^2+aa", ~2(a')^2+aa", ~2(a')^2+aa")\right],\\
R=R^\mu_{~\mu}=-6\left[\left(\frac{a'}{a}\right)^2+\frac{a"}{a}\right].
\end{eqnarray}
For general $a(t)$, the field equations (\ref{coupledRA}) take the form
\begin{eqnarray}
\nonumber
\partial_t^2A^0+3\frac{a'(t)}{a}\partial_t A^0-a^{-2}\sum_{k=1}^3 \partial_k\partial_k A^0+(6\xi+3\zeta)\frac{a"}{a}A^0\\+6\xi\left(\frac{a'}{a}\right)^2A^0=0,\\
\nonumber
\partial_t^2A^j+3\frac{a'(t)}{a}\partial_t A^j-a^{-2}\sum_{k=1}^3 \partial_k\partial_k A^j+(6\xi+\zeta)\frac{a"}{a}A^j\\+[6\xi+2\zeta]\left(\frac{a'}{a}\right)^2A^j=0,
\end{eqnarray}
for $j=1,2,3$. For the case of exponential expansion with present time at $t=0$, $a(t)=e^{\sqrt{\frac{\Lambda}{3}}t}$, $R_{00}=-\Lambda$, $R_{11}=R_{22}=R_{33}=\Lambda a(t)^2$ and $R=-4\Lambda$.\\

\noindent For the case of power-law expansion with present time at $t=1$, $a(t)=t^\alpha$,\\ $R_{00}=-3\alpha(\alpha-1)t^{-2}$, $R_{11}=R_{22}=R_{33}=\alpha(3\alpha-1)t^{2(\alpha-1)}$ and $R=6\alpha(1-2\alpha)t^{-2}$.

\subsection{Conformal time and Conformal Coupling}
Aiming to quantise the field equations, the coordinate system should be chosen so that the field equations are as close as possible to Lorentz-invariant wave equations for which the canonical quantisation method is well established. Therefore it is standard practice to adopt the conformal time coordinate $\eta$ with $t_0$ chosen so that the current time is either $t=1$ or $t=0$: 
\begin{equation}
\eta=\int_{t_0}^t \frac{1}{a(t_1)} dt_1.
\end{equation}
Then the metric conforms to that of Minkowski but with non-zero curvature invariants:
\begin{equation}
ds^2=a(t(\eta))^2\left(d\eta^2-(dx^1)^{2}-(dx^2)^{2}-(dx^3)^{2}\right).
\end{equation}
The leading cosmological paradigm \cite{Weinberg} is that of Lambda-Cold Dark Matter ($\Lambda CDM$). From soon after inflation, up to age 47,000 years, the energy was predominantly in the form of photon and neutrino radiation. During that period the dominant term in $a(t)$ was proportional to $t^{1/2}$. For most of the duration of the plasma universe which lasted up to the time of atomic recombination at age 360,000 years, the mass-energy was already dominated by matter. The time of last scattering of photons occurred around 380,000 years after the big bang. During the matter-dominated epoch, up to age 9.8 $\times 10^9$ years, the dominant term in $a(t)$ was proportional to $t^{2/3}$. Since then, the expansion has been predominantly exponential $a(t)\sim e^{t\sqrt{\Lambda/3}}$ where $\lambda$ is the cosmological constant, proportional to dark energy density.
Power-law and exponential expansion factors, converted to functions of $\eta$, are shown in \textbf{Table 1}.\\
\begin{table}[t]
	\caption{Expansion factor in terms of conformal time $\eta$. }
	{\begin{tabular*}{\textwidth}{@{\hspace*{5mm}}lll}
{a(t)} & {$\eta$(t) } & {a(t($\eta$)) }  \\\hline
			%& & \TCH{(\%)} &\\
			$(t+1)^\alpha$ ; $0<\alpha<1$ &  $\frac{(t+1)^{1-\alpha}-1}{1-\alpha}$ & \scalebox{0.9}{$\left[ 1+(1-\alpha)\eta\right]^{\alpha/(1-\alpha)}; \eta>\frac{1}{\alpha -1}$}\\\hline
			$t+1$ &  $\log (t+1)$ & $e^\eta$ \\\hline
			$(t+1)^\alpha$ ; $\alpha>1$ &  $\frac{(t+1)^{(1-\alpha)}-1}{1-\alpha} $& \scalebox{0.9}{$[1+(1-\alpha)\eta]^{\alpha/(1-\alpha)}; \eta<\frac{1}{\alpha-1}$}\\\hline
			$e^{t\sqrt{\Lambda/3} }$&  $\sqrt{\frac{3}{\Lambda}}\left[1-e^{-t\sqrt{\Lambda/3}} \right]$&$ \frac{\sqrt{3/\Lambda}}{{\sqrt\frac{3}{\Lambda}}-\eta}; \eta<\sqrt{3/\Lambda})$\\ \hline 
	\end{tabular*}}{
	}
	\end{table}
After changing time coordinate $t$ to $\eta$, the field equations are, with over-dot denoting $\eta$-derivative,
\begin{eqnarray}
\label{PDEetaA}
\partial_\eta^2 A^0-\sum_{k=1}^3\partial_k\partial_k A^0 +2\frac{\dot a(\eta)}{a}\partial_\eta A_0+{6\xi}\frac{\ddot a}{a}A^0
+3\zeta\left[\frac{\ddot a}{a}-\left(\frac{\dot a(\eta)}{a}\right)^2\right]A^0=0,\\
\partial_\eta^2 A^j-\sum_{k=1}^3\partial_k\partial_k A^j +2\frac{\dot a(\eta)}{a}\partial_\eta A_j+{6\xi}\frac{\ddot a}{a}A^j
+\zeta\left[\frac{\ddot a}{a}+\left(\frac{\dot a(\eta)}{a}\right)^2\right]A^j=0.
\end{eqnarray}

\subsubsection{Amplifying the signal}

In the partial differential equation for any of the components $A^\mu$, the first two terms are those of the standard d'Alembert wave equation. However the next term is a first-order damping term, again in common with each component equation. That term may be transformed away by changing the dependent variable to the amplified 4-potential $\bar A^\mu=a(t(\eta))A^\mu$. Note that $\bar A^\mu$ is still a \textit{bona fide} observable vector quantity. Its evolution equation still is derivable from a Lagrangian density and its quantum counterpart and the conjugate variables will still satisfy the CCR. The governing equations (\ref{PDEetaA}) are now expressed as 
\begin{eqnarray}
\label{PDEetaAbar}
 \partial_\eta^{~2} \bar A^0-\sum_{k=1}^3\partial_k\partial_k \bar A^0 +\left( [-1+6\xi+3\zeta]\frac{\ddot a}{a} -3\zeta \left[\frac{\dot a}{a}\right]^2\right)\bar A^0=0,\\
 \partial_\eta^{~2} \bar A^j-\sum_{k=1}^3\partial_k\partial_k \bar A^j +\left( [-1+6\xi+\zeta]\frac{\ddot a}{a} +\zeta \left[\frac{\dot a}{a}\right]^2\right)\bar A^j=0.
\end{eqnarray}
\subsubsection{Dynamical Permutability, Conformality and Instability.}

We say that the dynamical equations for 4-vector components $\bar A^0$ and $\bar A^j$ are \textit{permutable} if each component satisfies the same equation. Otherwise the components would be distinguishable by dynamical behaviour alone. Assume $a(\eta)$ is increasing and $a(0)=1$. From (\ref{PDEetaAbar}), dynamical permutability is equivalent to
\begin{eqnarray*}
\zeta\left[\left(\frac{\ddot a}{a}\right)-2\left(\frac{\dot a}{a}\right)^2\right]=0\\
\iff \zeta=0~~\hbox{or}~\left(\frac{\ddot a}{a}\right)=2\left(\frac{\dot a}{a}\right)^2\\
\iff \zeta=0 ~~\hbox{or}~a=\frac{\eta_\infty}{\eta_\infty-\eta},
\end{eqnarray*}
with $\eta_\infty$ constant.
Referring to {\bf Table 1}, we see that this dependence of $a$ on $\eta$ is equivalent to the expansion factor $a$ being exponential in cosmic time, with Hubble time $t_H=a/\dot a=\sqrt{3/\Lambda}=\eta_\infty$. Therefore we have the result:\smallskip 

{\textit{A dynamically permutable 4-vector potential can be coupled to the Ricci tensor}}\\ ($\zeta\ne 0$)
  {\textit{if and only if $a(t)$ is exponential in time}}.\smallskip
  
  Since the expansion factor approaches the exponential asymptotically, it may be considered that $\zeta$ is non-zero, and that the earlier radiation-dominated epoch and mass-dominated epoch allowed for relatively short-lived dynamically inconsistent stages in the long-term development of the universe. \\
 In dynamically permutable models, the vector field equation reduces to a Proca equation wherein each component satisfies the same Klein-Gordon equation with time-dependent squared mass:
 \begin{equation}
 \label{Proca}
  \partial_\eta^{~2} \bar A^\mu-\sum_{k=1}^3\partial_k\partial_k \bar A^\mu+M^2(\eta)\bar A^\mu=0. 
 \end{equation}
 Effectively, the electromagnetic field equations on a curved FLRW space-time have been transformed to a Proca vector equation on flat-space-time but with time-dependent squared mass.
  The case of conformal coupling is that which reduces the PDEs to the standard linear wave equation with zero mass. For non-exponential expansion, this requires $\zeta=0$ and $\xi=\frac 16$. This conformal value of $\xi$ is well known from scalar field theory.  For exponential expansion, the conformal condition is $\xi=\frac 16-\frac 14\zeta$, which includes the case $\xi=0$ and $\zeta=\frac 23$.\\
The following cases obey vector Proca-Klein-Gordon equations with positive time-dependent squared mass: (i) $a(t)=(1+t)^\alpha ; \alpha>\frac 12, \alpha\ne 1$, $\zeta=0$ and $\xi>\frac 16$,\\ (ii) $a(t)=(1+t)^\alpha ; 0<\alpha<\frac 12$, $\zeta=0$ and $\xi<\frac 16$, and\\ (iii) $a(t)=$ exponential and $\xi>\frac 16-\frac 14 \zeta$.\\ 
Cabral and Lobo derived a wave equation with $\zeta=1$ and $\xi=0$, which is dynamically stable with $a(t)$ exponential.\\
The less studied and less understood models are those such as the minimal coupling model ($\zeta=0, \xi=0$) in which $\xi<\frac 16-\frac 14 \zeta$, with power law $\alpha>\frac 12$ or exponential expansion so that $M^2(\eta)<0$. In those cases, the Hamiltonian is unbounded below, the classical solutions have some  non-oscillatory modes and some energy eigenstates of the quantum Hamiltonian are not eigenstates of a number operator that commutes with the Hamiltonian. That will be the case that we consider further here. The model with exponential $a(t)$ has \begin{equation}
M^2=[12\xi+3\zeta-2](\eta_\infty-\eta)^{-2},
\end{equation}
with $\eta_\infty=\sqrt{3/\Lambda}$. Note that with minimal coupling, the power law with very large $\alpha$ also is consistent with $M^2=-2(\eta_\infty -\eta)^{-2}$ which results from the exponential  $a(t)$. From the perspective of destabilisation, critical cases occur at $\alpha=0$ (static), $\alpha=\frac 12$ (radiation-dominated), $\alpha=1$ (Milne, equivalent to static) and $\alpha\to\infty$ ($\Lambda$-dominated). The case $\alpha=\frac 23$ (matter-dominated) has an additional special property that $\ddot a$ is constant.\\ 
\section{Field Quantization}
From our first introduction to dynamics, we learn that models may become mathematically more tractable when we introduce additional variables that have some redundancy. The mechanics of a particle in a conservative force field is conceptually simpler after we introduce a potential function that is not uniquely defined but allows us to construct a first integral, the energy.  This idea extends to the electromagnetic 4-potential. It allows us to collapse Maxwell's equations to a simpler mathematical form in terms of a set of potential variables with some redundancy. In Minkowski space, the dynamical equation  for the 4-potential has gauge invariance $A_\mu\to A_\mu+\psi,_\mu.$ A specific choice of gauge can be designed to facilitate a specific purpose, such as reduction to a standard wave equation which follows directly in the Lorenz gauge (\ref{Lorenz}). At the same time, a specific choice of gauge removes some of the redundancy. Even after choosing a gauge, it first appears that there are three degrees of freedom at each point in space-time, when we know that there are two independent physical polarisations of electromagnetic waves. The extra redundancy is associated with the fact that the Lagrangian density (\ref{usualL}), whether or not the Proca mass term is included, does not depend explicitly on $A^0,_0$.  A conjugate momentum variable $\Pi_{0}$ cannot be defined. Consequently, canonical quantisation cannot proceed in the normal way. These comments are made in view of the historical context of extending  the Proca system to a gauge theory, which eventually led to first-class constraints, a unitary representation of the commutation algebra on a positive normed space, and  to renormalizability.\\
Before we consider quantising the non-autonomous Proca system (\ref{Proca}), as a starting point, we can refer to the long and complicated history of the quantised Proca field with constant positive mass.  
\subsection{Standard Autonomous Proca Field}
The standard Proca equation describes a vector Boson field with constant mass, in which case the components $A^\mu$ are not  potentials but physical fields. The Proca equation originally appeared in the form 
\begin{equation}
\label{Proca1}
\partial_\nu F^{\nu\mu}+m^2A^\mu=0.
\end{equation}
By taking $\partial_\mu$ of each side, by the skew-symmetry of $F^{\nu\mu}$, the first term is annulled and we are left with the Lorenz condition (e.g. reference \cite{Ruegg}). The Proca equation is not gauge-invariant in the sense of electromagnetism. The Lorenz condition still must be imposed as a constraint. This is a troublesome second class constraint \cite{Dirac}, formally $\mathcal W[{\bf A}]=0$, although $\mathcal W$ does not commute through Poisson brackets with a naively constructed Hamiltonian.  After quantisation it is not clear how a complete set of physical eigenstates of a Hamiltonian could belong to the zero-eigenspace of $\mathcal W$ and of its Hermitian conjugate  as an operator in the Hilbert space representation. Since the earliest formulations of quantum vector field theory, a large body of work in mathematical physics has been devoted to various pathways to quantise systems with second class constraints. In the language of algebraic quantisation, this may be cast as the existence of regular states on the CCR  algebra of observables that incorporates supplementary conditions \cite{Grundling}. In limited space, it would be difficult to do justice to a literature survey of the various concrete approaches that have been developed. They are described in the comprehensive review in reference \cite{Zamani}. The two most common approaches can be classified as (i) extending classical Poisson brackets to Dirac's constraint brackets before quantisation \cite{Dirac}, and (ii) introducing one or more new fields that allow a new choice of gauge to change the second class constraints to first class constraints. The additional St\"uckelberg field $\rho$ was introduced in 1938 \cite{Stuck} in the Lagrangian:\\
\begin{equation}
\label{StuckL}
\mathcal L=-\frac 12 \partial_\mu A^\nu\partial^\mu A_\nu +\frac 12 m^2A^\mu A_\mu+\frac 12\partial_\mu\rho\partial^\mu\rho-\frac 12 m^2\rho^2.
\end{equation}
 This allowed a new type of gauge invariance \cite{Pauli}:
\begin{eqnarray}
A_\mu\mapsto A_\mu +\partial_\mu\chi,\\
\rho\to \rho+m\chi,\\
\partial_\mu\partial^\mu\rho+m^2\rho=0,\\
\partial_\mu\partial^\mu\xi+m^2\chi=0.
\end{eqnarray}
Instead of the Lorenz condition, one may impose a first-class constraint that involves all of the field variables.
$\mathcal L$ now depends on $\partial_0A^0$ so a conjugate momentum variable has the familiar definition $\Pi^\mu=\partial_0\partial\mathcal L/ \partial A^\mu_{,0}.$ A Hamiltonian can be constructed. Integrating over a spatial section $\Omega$, 
$$H=\frac 12\int_\Omega -\left[(\Pi^0)^2+m^2(A^0)^2 +\nabla A^0\cdot \nabla A^0\right]+ \sum_{j=1}^3(\Pi^j)^2+m^2(A^j)^2 +\nabla A^j\cdot \nabla A^j$$
\begin{equation}
+\left[(\Pi^5)^2+m^2\rho^2 +\nabla \rho\cdot \nabla \rho\right]dV,
\end{equation}
where $\Pi^5$ is the momentum variable conjugate to $\rho$.
The first part with index 0, is negative and unbounded. However, as in the Gupta-Bleuler quantisation of the electromagnetic field, this was associated with unphysical states of negative norm, dynamically unconnected to physical states of positive norm. In general, Lorentz covariant locally gauge invariant fields must be represented on an indefinite inner product space \cite{Strocchi}. Note that the matrix elements of the constraint operators showed zero mappings from the physical subspace to the unphysical subspace. Importantly, in the St\"uckelberg-Feynman gauge this then led to the same equal-time commutation relations among $A^\mu(x)$ and $\Pi^\mu(y)$ as in standard elecftrodynamics. With a non-zero time difference between operators, the commutators involved one solution of the massive Klein-Gordon equation.\\
The historical extensions of St\"uckelberg's formulation and its more recent ramifications, are described in reference \cite{Ruegg}. Neglecting an additional inconsequential term that is a total derivative, the Lagrangian may be re-expressed
\begin{equation}
\label{Lagrangian2}
\mathcal L=-\frac 14 F_{\mu\nu}F^{\mu\nu}  +\frac 12 m^2(A_\mu-\frac 1m\partial_\mu \rho)(A^\mu -\frac 1m\partial^\mu \rho)-\frac 12(\partial_\mu A^\mu +m\rho)^2.
\end{equation}
The mass parameter is seen to be a coupling constant between the physical field and the gauge field. In this regard, St\"uckelberg's construction is seen to be a fore-runner to the Higgs mechanism. In the Feynman-St\"uckelberg gauge, the last term can be equated to zero at the end of calculations. This allowed Feynman to quantise the field theory via path summations. Renormalisability of more general fields eventually followed through appropriate gauge fixing after the introduction of one extra real parameter along with a Faddeev-Popov Fermionic ghost field \cite{Delbourgo}. \\
\subsection{Proca Equation with Time-dependent Mass}

We are particularly interested in the case that the mass parameter in (\ref{Proca}) is negative and decreasing in time. This includes the case of minimal coupling.  (\ref{Proca}) may still be derived from a real-valued Lagrangian density that is (\ref{StuckL}) with $A^\mu$ replaced by $\bar A^\mu$, $m^2$ replaced by $M^2(\eta)$ and $\rho$ set to zero. In (\ref{Lagrangian2}), the second term is no longer invariant under the Pauli gauge symmetry since a change of gauge introduces unbalanced terms that are quadratic in $\dot M$. At this stage, a gauge theory with time-dependent mass has not been developed. 
However the constraints and gauge conditions are treated, it remains true that each component $A^\mu$ satisfies a Klein-Gordon equation with negative mass that is decreasing in time. In the case of a spatially flat universe that is expanding exponentially or by power-law in cosmic time, $M^2$ is proportional to $-(\eta_\infty- \eta)^{-2}$. In those cases, with homogeneous Dirichlet or Neumann boundary conditions, the complete set of classical solutions of the non-autonomous Klein-Gordon equation can be constructed by separation of variables \cite{Deutscher}.  In terms of spherical harmonic functions of spherical polar coordinates and Bessel functions of $r$,
\begin{equation}\label{phieq}
\bar A^\mu=
\sum_{n=1}^\infty\sum_{l=0}^{2n}\sum_{m=-l}^l
\frac{J_{l+\frac12}(k_{nl}~r)}{\sqrt{k_{nl}~r}}   Y_l^m(\theta,\varphi)   \{{a^\mu}_{nlm}f_{n\ell}(\eta)+(-1)^m {a^\mu}_{nl-m}^\dagger f_{n\ell}^*(\eta)\}.
\end{equation}
The time dependent factors satisfy the equation of an oscillator with time dependent angular frequency $\omega_k=\sqrt{k^2-12\gamma(\eta_\infty-\eta)^{-2}}$:
\begin{equation}
\ddot f_k+\left[k^2-12\gamma(\eta_\infty-\eta)^{-2}\right]f_k=0;~~\gamma=-\xi+\frac 16(1-\zeta).
\end{equation}
The solutions are linear combinations of Bessel functions

\medskip

$(\eta_\infty-\eta)^{1/2}J_{\sqrt{12\gamma+1/4}}(k[\eta_\infty-\eta])~~~\hbox{and}~~(\eta_\infty-\eta)^{1/2}Y_{\sqrt{12\gamma+1/4}}(k[\eta_\infty-\eta]).$
\medskip

Bessel functions of the second kind are ruled out because of their singularity at $\eta=\eta_\infty$. Then the solution reaches its last extremum at time $\eta=\eta_\infty-\lambda'_{\nu,0}/k_{n\ell}$, where $\lambda'_{\nu,0}$ is the smallest positive local extremum point of the Bessel function of order $\nu=\sqrt {12\gamma+1/4}$. The subsequent period of time up to $\eta=\eta_\infty$ is of infinite duration in terms of cosmic time.\\
The un-amplified signal is $A^\mu=\bar A^\mu /a(\eta)$. When the order of the Bessel function is real, $A^\mu$ has temporal amplitude of the order $(\eta_\infty-\eta)^\kappa$ with $\kappa=\frac 32-\sqrt{\frac 94-2\zeta+12\xi}$. This amplitude is constant in the case of minimal coupling, or more generally when $\zeta+6\xi=0$. \\Here, we have assumed a domain that is not elliptoid but spherical $r\le 1$.  This applies in a reference frame in which the cosmic background radiation is isotropic. Anisotropic redshifts show that our galaxy is moving at approximate speed 670 km $s^{-1}$ relative to that frame \cite{Consoli}. 
Unlike in infinite Minkowski space, the wave numbers $k$ take a countable spectrum of values $k_{n\ell}$. In unstable cases such as minimal coupling, each angular frequency decreases in time until it becomes zero and then purely imaginary. These attraction-repulsion transitions take place at later times for successively higher wave numbers. Each mode oscillates only a finite number of times in the future beyond any finite starting time.\\
When the solutions are known, calculation of the magnitude squared gradients in the Hamiltonian can be considerably simplified.
Green's identity establishes
$$\label{greens}
\int_\Omega\phi\nabla^2\psi+\nabla\phi\cdot\nabla\psi \,dV=\int_{\partial\Omega}\phi\, \hat n \cdot\nabla\psi\, dS
$$
where $\partial\Omega$ is the spatial boundary. Then with $\psi=\phi$, if we consider homogeneous boundary conditions of either  Dirichlet or Neumann type, the boundary integral vanishes. Consequently
\begin{equation}
\nonumber
\frac 12\int_\Omega (A^\mu_{,\eta})^2 +\nabla A^\mu\cdot \nabla A^\mu+M^2 (A^\mu)^2 ~dV= \frac12\int_\Omega( A^\mu_{,\eta})^2-A^\mu A^\mu_{,\eta\eta}\,dV
\end{equation}
In terms of material coordinates, we prefer the Neumann condition because in the usual physical coordinates, the stress-energy tensor then indicates zero transport of 4-momentum across the moving fluid material boundary. That boundary condition leads to the eigenvalues $k_{n\ell}$ being zeros of derivative spherical Bessel function $j'_\ell$ for any $\ell=0,1,2,\cdots$. With Dirichlet boundary conditions, they would be zeros of $j_\ell$. The spacing and interlacing of those eigenvalues are similar in the two cases. At large n, $k_{n\ell}\sim N\pi$.\\

A direct calculation \cite{Deutscher} shows that the Hamiltonian at each frozen time is the sum of a repulsive component with finite but increasing degrees of freedom, plus the usual oscillatory component with infinite degrees of freedom. The number of repulsive modes is the number of independent modes that have pure imaginary angular frequencies. Due to the simplicity of the equation for $f_k(\eta)$, at each frozen time an unstable mode has the simple quadratic Hamiltonian of an oscillator with imaginary frequency, unbounded below, in one degree of freedom. \\

In principle, at each time, the Hamiltonian of the oscillatory part could be quantised in the same way as for the standard Proca system. At each wave number, gauge fixing effectively reduces the number of independent  oscillatory modes to that of the usual angular momentum labels $\ell$ and $m$  multiplied by three, allowing for two  independent  transverse oscillations plus one longitudinal oscillation. Longitudinal oscillations are allowed in an expanding space because the expansion brings an effective mass to the equivalent Proca field. However, in the unstable modes, the word ``oscillation" does not apply.\\

For the non-oscillatory component, there is an unstable runaway. States of negative energy are unavoidable and they are not isolated by the dynamics. At each frozen time, the non-oscillatory part of the Hamiltonian has a continuous spectrum that is unbounded below. That is the same phenomenon of ``jelly states" \cite{Schroer} in the case of M being negative and constant. The extra complication here is that $M(\eta)$ is decreasing. The level of degeneracy at each energy level $n$, increases as $n^2$.  Consequently, the density of modes per unit volume, increases as $n^2$ \cite{Kodama}. The critical value n=N at the onset of instability, increases asymptotically in proportion to expansion factor $a(\eta)$. Therefore the total number of unstable modes increases asymptotically as $a(\eta)^3$ (\cite{Deutscher}). Assuming a sharp explosive initial condition with broad spectrum, this accounting is in accord with the energy density of dark jelly being constant. In dimensionless terms, under the exponential expansion with $a=e^{\sqrt{\Lambda/3}t}$, a wave number becomes critical when 
\begin{equation}
k_{N,\ell}=\left(\frac 23 \Lambda\right)^{1/2}a(t)~~\sim N\pi a(t)~~\hbox{for N large.}
\end{equation}
At the present time with $a=1$, in terms of dimensional quantities, the length scale is $R_0$, which is thought to be much larger than the observable distance to the horizon of the universe. Within errors of measurement, the spatial curvature out to the horizon is zero. This uniform flatness has been explained by the inflation postulate. From the low curvature, the universe must be at least 250 times larger than the observable universe, which itself is 3.3 times larger than the Hubble radius $ct_H$. It is a coincidence that the Hubble time of 14 billion years is close to the age of the universe at 13.7 billion years. Through exponential expansion, $t_H$ will remain constant as the age increases. Now the critical wave number has
\begin{equation}
N\approx \frac{\sqrt 2 R_0}{\pi ct_H}> 380.
\end{equation}
Assuming that all of the independent unstable modes have evolved from independent stable modes, the number of independent unstable modes must be greater than $3\times 380^3>1.6\times 10^8$. 

\section{Conclusion} 
We have developed a generally covariant vector equation for the long-term evolution of the electromagnetic field. Such a general covariant evolution is governed by a Lagrangian density that has coupling with the Ricci curvature scalar as well as with the Ricci tensor. However our analysis includes the important case of minimal coupling in which the curvature is represented only in the Laplace-Beltrami operator that generates the flow, with both external coupling constants set to zero. Within that scheme, each choice of values for the two coupling constants defines one model. For an important class of models, the vector equation may be transformed, by a change of variables to a Proca equation with time dependent decreasing mass. With effective non-zero mass, we need to accept that universal stretching allows for the possibility of longitudinal oscillations, at least over a very long time scale.\\

In the long term, continued exponential expansion of the universe will surely destabilise electromagnetic fields. Successively at shorter material wavelengths, each independent dynamical mode will become repulsive at some discrete time.  Beyond the present time, any dynamical mode will have only a finite number of oscillations. At any time there will be only a finite but increasing number of unstable modes, governed by a Hamiltonian in finite degrees of freedom. A dynamical mode becomes unstable only after its wavelength has been stretched by cosmological expansion, to a length that is comparable to the Hubble radius. These modes will show little spatial variation over a scale that is enormously large compared to our field of perception. However the observable universe is actually small compared to the radius of the universe beyond the horizon, that is needed in the inflationary models to explain the uniformly high degree of spatial flatness that is observed as far back into the distant past as we can see. That means that even by the current time, more than 160 billion independent modes must have become unstable. They would have little or no discernible spatial structure. Their energy would be diffuse over a very long scale, with negligible energy contained in a region that is small enough to react with laboratory-scale equipment. \\
 
 Although this work is by no means complete, we may infer some properties of the quantised model from established results on quantum field theory of the Proca equation with constant mass.   In keeping with the wave-particle duality, the absence of wavelike behaviour is equivalent to the absence of particle-like behaviour. At any particular time the unstable modes are governed by a quantum Hamiltonian that does not commute with any number operator. We cannot count particles in eigenstates of the time-frozen Hamiltonian. There is no state of lowest energy and there is no normalisable state that has the usual invariance properties of a vacuum. Unlike in the Gupta-Bleuler approach to the electromagnetic field, we cannot define a separate physical subspace from which the dynamics do not transgress to the unphysical sector that requires states to have negative energy and negative norm. In our experience, these phenomena of unstable quantum systems are not widely studied and not widely understood. Quantum many-body theory has been informed by knowledge of stable oscillatory systems. Fortunately the unstable component of a field coupled to expanding space-time, has only finite degrees of freedom. In order to better understand some of these properties of unstable quantum systems, Section 2 has drawn together some different approaches to the theory of quadratic Hamiltonians in quantum mechanics. The group of Bogoliubov transformations is equivalent to the symplectic group of linear canonical transformations on phase space. The angular frequencies of a linear system are invariant under those transformation groups. Therefore it is easily seen that an unstable system with complex frequencies cannot be transformed to a stable system for which the Hamiltonian is a linear combination of number operators plus a constant vacuum energy. Therefore the quantum Hamiltonian of an irreducible unstable system cannot commute with a number operator. The particle interpretation breaks down. The quantum Hamiltonian has continuous spectrum, unbounded below.\\
 
 For the unstable component of the field with finite degrees of freedom, the Stone- von Neumann theorem tells us that any irreducible  representation of the quantum algebra of observables as self-adjoint operators, must be unitarily equivalent to any other. The choice of representation of that component will make no practical difference. However for the stable part of the field, which has infinite degrees of freedom, the Fock representation is still available, out of an uncountable infinity of inequivalent representations.\\

\textit{Acknowledgments}\\
PB is grateful to the Australian Research Council for funding of Discovery Project DP220101680.
	
	% Author details will always appear the end of the chapter in the final version of the chapter
	
	\textit{Copyright:}The above-named authors hold copyright on this article. Quoting any part of it or copying any part in a publication, requires attribution of the authors.
	% Note: The copyright year will be changed accordingly during production to correspond with the year of publication.

\end{document}